\documentclass[aps,twocolumn,showkeys,unsortedaddress]{revtex4-1}
\usepackage{graphicx}
\usepackage{amsmath}
\usepackage{amssymb}
\usepackage{longtable}

\usepackage{natbib}
\setcitestyle{round}
\begin{document}

\title{Conditional solvation of isoleucine in model extended and helical peptides: context dependence of hydrophobic hydration and limitations of  the 
group-transfer model}
\author{Dheeraj S. Tomar}
 \affiliation{Department of Chemical and Biomolecular Engineering, Johns Hopkins University, Baltimore, MD}
\author{Val{\'e}ry Weber}
 \affiliation{IBM Research, Zurich, Switzerland}
 \author{B.~Montgomery Pettitt}
 \affiliation{Sealy Center for Structural Biology and Molecular Biophysics, University of Texas Medical Branch, Galveston, TX}
\author{D. Asthagiri}\email{Corresponding author: Dilip.Asthagiri@utmb.edu}
\affiliation{Department of Chemical and Biomolecular Engineering, Johns Hopkins University, Baltimore, MD}
 \affiliation{Sealy Center for Structural Biology and Molecular Biophysics, University of Texas Medical Branch, Galveston, TX}

 \begin{abstract}
The hydration thermodynamics of  the GXG tripeptide relative to the reference GGG is often used to define the \textit{conditional} hydration contribution of X. This quantity or the hydration thermodynamics of a small molecule analog of the side-chain or some combination of such estimates, usually including factors to account for the solvent-exposure of the side-chain in the protein,  have anchored the interpretation of seminal experimental studies in understanding protein stability and folding.  Using simulations we show that such procedures to model protein hydration thermodynamics have significant limitations. We study the conditional hydration thermodynamics of the isoleucine side-chain in an extended pentapeptide and in helical deca-peptides
using corresponding extended penta-glycine or helical deca-peptides as reference. Hydration of butane in the \textit{gauche} conformation provides a small molecule reference for the side-chain. We use the quasichemical approach to parse the hydration thermodynamics into chemical, packing, and long-range interaction contributions. The chemical contribution of $g$-butane,
reflecting the contribution of solvent clustering within the defined inner-shell of the solute, is substantially more negative than the conditional chemical contribution of isoleucine. The packing contribution gives the work required to create a cavity in the solvent,  a quantity of central interest in understanding hydrophobic hydration. The packing contribution for $g$-butane substantially  overestimates the conditional packing of isoleucine. The net of such compensating contributions  disagrees with the conditional free energy of isoleucine but by a lesser magnitude.  The excess enthalpy and entropy of hydration of $g$-butane model are also more negative than the corresponding conditional quantities for the side-chain.   The conditional solvation of isoleucine in GGIGG also proves unsatisfactory in describing the conditional solvation of isoleucine in the helical peptides.
\end{abstract}

 \maketitle

\section{Introduction}

The hydration thermodynamics of analogs of amino acid side-chains or of amino acid side-chains in small model peptides 
have often been used to understand protein folding and protein-protein association. Examples of such approaches abound in the biochemical literature. An admittedly biased list of a very small number of the many pioneering investigations where this approach has been used include work identifying hydrophobicity as a dominant force in protein folding \cite{kauzmann:59}, 
 investigations of protein denaturation \cite{Tanford:62,Nozaki:1963,Tanford:1964,Tanford:apc70},   interpretation of calorimetric data on protein unfolding \cite{makhatadze:jmb1993,privalov:jmb1993,makhatadze:1995}, and more recently, investigations on the role of osmolytes in protein folding \cite{Auton:bc11,Auton:pnas07,Auton:pnas05}.  Drawing upon insights attributed to Cohn and Edsall,  Tanford \cite{Tanford:62,Tanford:1964} formulated this approach into a quantitative, predictive framework. In his approach,  the free energy of unfolding is given as a sum of the free energy of transfer of ``the small component groups of the molecule, from the environment they have in the native form, to the environment they have in the unfolded form'' \cite{Tanford:62}.  Accounting for subsequent refinements that included corrections for the solvent-exposure of 
the ``small component groups'' (cf.\ Ref.~\onlinecite{Auton:bc11}), the generic equation of the unfolding free energy ($\Delta G_{N\rightarrow U}$) can be written as 
\begin{equation}
\Delta G_{N\rightarrow U} = \sum_i \alpha_i \Delta g_i \, ,
\end{equation}
where $\Delta g_i$ is the free energy of transferring the group $i$ from some reference phase to liquid water, and $\alpha_i$ is a factor that corrects for the change in solvent-exposure of the group $i$ in protein unfolding. In interpretation of calorimetric
data \cite{makhatadze:1995}, for example,  a gas-phase reference is natural, as is also the case in this article. The same form of equation can also be used to describe the effect of osmolytes on protein unfolding: identifying $\Delta g_i$ with the free energy for transferring the group between water and the aqueous osmolyte solution and  $\alpha_i$ with the change in the solvent exposure upon unfolding in the osmolyte solution relative to water,  the above equation directly gives the so-called $m$-value for 1~M osmolyte \cite{pace:jbc74}. 

At a time when theory, simulations, and experimental techniques were much less developed than they are now, the group-additive approach a pragmatic first step to understand the hydration thermodynamics of a complicated macromolecule. But it is also important to assess its limitations, and probe if the \textit{physical conclusions\/} based on this approach are valid.  For example, the group-additive model appears to capture the effect of osmolytes rather well \cite{Auton:bc11}: the 
predicted and experimentally determined $m$-values agree to within a couple kcal/mol for proteins with
about 100 residues. On this basis it has been suggested that osmolytes largely act by tuning the solubility of the peptide
backbone \cite{Bolen:2008bf}. Our recent simulation-based scrutiny of the transfer for a peptide, however, suggests that the
good agreement in $m$-values likely arises because solvent-mediated correlations between two groups largely cancel in the water-to-osmolyte transfer \cite{tomar:bj2013}. Further, in the vapor-to-liquid transfer, a situation where solvent-mediated correlations are preserved,  the identified group-additive contribution of the peptide group depends
on whether blocked-Gly$_n$ or cyclic-diglycine was used as a model: that is the group-contribution was not
context independent as often assumed.

Building on the work noted above \cite{tomar:bj2013}, and other recent studies that have exposed
limitations of the additive model \cite{helms:2005fw,Boresch:jpcb09,Boresch:bj13},  here we study how well a model of the side-chain describes the hydration of the side-chain and, most importantly, whether the physical conclusions about hydrophobicity of a model are applicable to the side-chain. The problem we consider is the  vapor-to-water transfer (hydration) of an isoleucine side-chain in the context of model extended and helical peptides. Butane in the \textit{gauche} conformation, matching exactly the side-chain conformation of the isoleucine in extended peptides, is used as a small-molecule analog of the isoleucine side-chain. 

As before, we parse the hydration free energy into hydrophobic and hydrophilic contributions. We find that the net hydration
free energy of the target peptide IGGGG or GGIGG is predicted to within 2~kcal/mol (in a net free energy of about 30~kcal/mol) using the hydration free energy of GGGGG and $g$-butane, appropriately adjusting for solvent exposure of GGGGG and 
isoleucine side-chain in the target peptide. But,  the underlying components convey a different picture.  Using the $g$-butane model will cause one to overestimate substantially the hydrophobicity of the isoleucine side-chain in the peptide models. Since the $g$-butane model also underestimates the hydrophilic contributions, the net chemical potential works out reasonably well.   Importantly, using group-additivity the magnitude of the excess enthalpy and entropy of hydration are also incorrectly predicted, but these also balance one-another. 
Similar trends are also found when we use the conditional hydration free energy of isoleucine in GGIGG to predict the
hydration thermodynamics of model helical deca-peptides. 

Our work suggests that while the group-additive approach may describe the net free energy reasonably well, 
caution is necessary in drawing physical conclusions based on this observed agreement. 

\section{Theory}

The excess chemical potential $\mu^{\rm ex}$ and its enthalpic ($h^{\rm ex}$) and entropic ($s^{\rm ex}$) components are 
of primer concern in this work. The excess chemical potential is that part of the free energy that would vanish if all intermolecular interactions were neglected. Formally, $\mu^{\rm ex} = \langle e^{\beta \varepsilon}\rangle$,
where it is understood that $\mu^{\rm ex}$ is referenced relative to the ideal gas at the same density and temperature. The averaging $\langle \ldots\rangle$ is over the solute-solvent binding energy
distribution $P(\varepsilon)$ and  $\beta = 1/k_{\rm B}T$, with $T$ the temperature and $k_{\rm B}$ the Boltzmann constant. 

The direct approach to calculate $\mu^{\rm ex}$ by characterizing $P(\varepsilon)$ usually fails since the high-energy
(high-$\varepsilon$) regions of $P(\varepsilon)$ are never well-sampled in simulations. The usual approach 
then is to calculate $\mu^{\rm ex}$ by accumulating the work done in changing the strength of the solute-solvent interaction from a non-interacting solute to the fully-interacting solute. Here we approach the problem differently, with a view to better
dissecting the physics of hydration. 

Imagine an inner-shell (or hydration shell), defined by a length parameter $\lambda$, around the solute.   At equilibrium, the solute samples a distribution of coordination states $\{n\}$, where $n$ is the number of water molecules within the inner shell. The association of the $n$-water molecules with the solute can be described in terms of an equilibrium constant $K_n$. 
A basic mass-balance then gives the probability $x_0$ that the inner-shell is empty of solvent by $\ln x_0 = -\ln \left(1 + \sum_{i\geq1} K_i \rho_w^i\right)$, where $\rho_w$ is the bulk density of water \cite{lrp:apc02,lrp:book,lrp:cpms,merchant:jcp09}.  The free energy to allow water molecules to populate a formerly empty inner-shell is just $k_{\rm B}T\ln x_0$. For a solute with an empty inner-shell the solute-solvent interactions
are necessarily of a longer-range than for the solute without this restriction. But these long-range interactions are better behaved,
such that for a sufficiently large inner-shell (typically up to the first hydration shell), $P(\varepsilon | \phi_\lambda)$, the probability density of the solute-solvent binding energy with the restriction of an empty inner-shell (denoted by $\phi_\lambda$), is Gaussian. 
Thus the specification of the inner-shell helps separate strong, short-range solute-water interactions from weaker, non-specific, longer-range interactions. 

We can consider an analogous process of solvent clustering within the same defined inner-shell but in the absence of the solute;
$p_0$, the probability to find a cavity in the solvent, has an expansion similar to $x_0$.  The quantity $-k_{\rm B}T\ln p_0$ is the work done to create a cavity in the liquid and is of principal interest in understanding hydrophobic hydration \cite{Pratt:1992p3019,Pratt:2002p3001}. 

\begin{figure}[h!]
\centering
\includegraphics[width=3.25in]{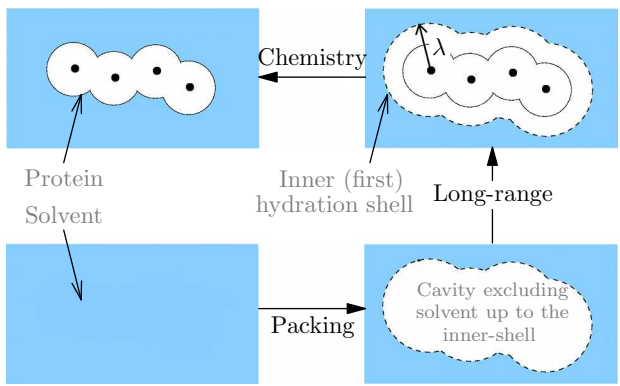}
\caption{Schematic depicting the quasichemical organization, Eq.~\ref{eq:qc}). Schematic reproduced from Ref.~\onlinecite{tomar:bj2013} with permission
from Elsevier.}\label{fg:cycle}
\end{figure}

In terms of the short-range chemical, long-range nonspecific, and cavity (packing) contributions, 
we then have \cite{lrp:apc02,lrp:book,lrp:cpms,Weber:jctc12,tomar:bj2013}
\begin{eqnarray}
\beta\mu^{\rm ex} = \underbrace{\ln x_0[\phi_\lambda]}_{\rm local\; chemistry} \underbrace{- \ln p_0[\phi_\lambda]}_{\rm packing} + \underbrace{\beta\mu^{\rm ex}[P(\varepsilon|\phi_\lambda)]}_{\rm long-range\; or\; outer} \; .
\label{eq:qc}
\end{eqnarray}
Figure~\ref{fg:cycle} provides a schematic of the decomposition of $\mu^{\rm ex}$ according to 
Eq.~\ref{eq:qc}. The constraint $\phi_\lambda$ is essentially an a field of range $\lambda$ that is used to regularize the problem of calculating $\mu^{\rm ex}$ from $P(\varepsilon)$. 
In simulations, we grow the external field $\phi_\lambda$ (instead of changing the underlying system Hamiltonian) to calculate $x_0$ and $p_0$ \cite{Weber:jctc12,tomar:bj2013}. In practice, the field has a soft-boundary at $\lambda$ \cite{Weber:jctc12}, unlike the sharp-demarcation suggested in the schematic. It is straightforward to correct for this effect \cite{weber:jcp11}, but we do not pursue that here. 

For $P(\varepsilon|\phi_\lambda)$ a Gaussian, we have 
\begin{eqnarray}
\mu^{\rm ex}[P(\varepsilon|\phi_\lambda)] = \langle \varepsilon | \phi_\lambda\rangle + \frac{\beta}{2} \langle \delta\varepsilon^2 | \phi_\lambda\rangle \, .
\label{eq:gaussian}
\end{eqnarray}

The excess entropy of hydration of the solute is given by 
\begin{eqnarray}
	T s^{\rm ex} & \approx & E^{\rm ex} - \mu^{\rm ex}
\label{eq:entropy}
\end{eqnarray}
where $E^{\rm ex}$, the excess energy of hydration, is essentially equal to the excess enthalpy ($h^{\rm ex}$) of hydration. The above equation ignores small corrections due to a finite excess volume of hydration and corrections arising from the thermal expansion coefficient and the isothermal compressibility of the pure liquid.

\subsection{Correlation effects in hydration}\label{sec:qual}

The above framework is particularly apposite to investigate the additive description of $\mu^{\rm ex}$.  To this end, for simplicity consider the hydration of a dimer AA. Within the additive model the packing contribution of AA is obtained from the packing contribution of A alone. Given that a cavity is already formed around one center, the work required to create the cavity around the second center will necessarily be less than  the work required to form the same cavity in isolation. This is because for the cavity in isolation more water molecules are pushed away and a larger interface is created than when another cavity is in proximity. Thus a naive additive approximation of the cavity contribution for the pair AA will be more positive than dictated by the molecular
reality.  For the same reasons, less work is needed to empty the inner-shell around the pair AA than given by the sum of the chemistry contribution for each A. That is, the pair additive description of the chemistry contribution will always be more negative than dictated by molecular reality. 

Such correlation effects are also present in the long-range contribution \cite{tomar:bj2013,paulaitis:corr10}. In this case, 
however, the correlated fluctuation of binding energies can either raise or lower the free energy. For AA, we expect
the favorable interaction of the solvent with one center to come at the expense of favorable interaction with the other center. 
For such anti-correlated binding energies, the free energy of the pair is lowered, as has been discussed earlier \cite{tomar:bj2013,paulaitis:corr10}. Even simple glycyl-peptides show context-dependent positive and negative correlations.

In  practice, the $\mu^{\rm ex}$ of A is usually scaled by extent of solvent exposure in AA to model the $\mu^{\rm ex}$ of the 
dimer (see Eq.~\ref{eq:areascaling} below). In terms of the quasichemical decomposition, we can appreciate that the area-scaling tries to correct for  the overestimation of the packing contribution and the underestimation of the chemistry contribution. How a
a local area-based scaling can correct long-range interaction contributions is difficult to appreciate, but
this is precisely what is done in experimental analysis.

\section{Methods}

The pentapeptides GGGGG, GGIGG, and IGGGG are modeled in the extended configuration with the long axis aligned with the diagonal of 
the simulation cell and the center of the peptide placed at the center of the simulation cell.  Helical deca-glycine (G$_9$G) and a helical peptide with
nine alanine and one glycine residues (A$_9$G) served as the reference for the helical peptides, G$_9$I and A$_9$I, respectively. In the helical peptides, isoleucine was substituted at position 6, roughly in the center of the peptide. Butane in the \textit{gauche} conformation was built using the isoleucine conformation in the extended pentapeptide.

The G$_9$G and A$_9$G peptides were built from a helical deca-alanine. These structures
were extensively energy minimized with weak restraints on heavy atoms to prevent fraying of the helix. The G$_9$I and A$_9$I helices were
built by appropriately grafting the conformation of isoleucine in the GGIGG system onto position 6. (Thus we ensure that the internal conformation of the
isoleucine is the same in both $g$-butane, GGIGG, and the helical peptides.) All peptides had an acetylated (ACE) N-terminus and an n-methyl-amide 
(NME) C-terminus. 

After energy minimization, the peptide atoms are held fixed throughout the subsequent simulation. The solvent was modeled by the TIP3P \cite{tip32,tip3mod} model and the CHARMM \cite{charmm} forcefield with correction terms for dihedral angles \cite{cmap2}  was used for the peptide.  A total of 2006 TIP3P molecules solvated the 
pentapeptide; 3500 water molecules were used for studies with helical peptides. 

The conditional hydration free energy of isoleucine in GGIGG relative to the model GGGGG is defined by $\mu^{\rm ex}[{\rm I} | {\rm GGIGG}] = \mu^{\rm ex}[{\rm GGIGG}] - \mu^{\rm ex}[{\rm GGGGG}]$ \cite{BenNaimVol2}. Similar definitions apply to the excess enthalpy and entropy of hydration and to isoleucine in context of other reference systems.  In experiments it is often the case that the hydration thermodynamics of the GXG tripeptide is compared with the GGG peptide to ascertain the contribution of X alone. The zwitterionic amino acid X alone is not used primarily to minimize end-effects. By this yardstick, we should expect the GGIGG peptide constrained to be in the extended conformation to be an even more conservative model to ascertain the contribution of the isoleucine side-chain. Capping the ends with ACE and NME further serves to dampen end-effects arising from long-range electrostatic interactions.

The free energy calculations and error analysis exactly follow the procedure described earlier \cite{tomar:bj2013}. Briefly,
the free energy to apply the field is obtained by growing the field between 0 and 5~{\AA}. We use Gauss-Legendre quadratures with  five gauss-points per unit {\AA}ngstrom to integrate the force exerted by the field on the solvent molecules versus the distance $\lambda$. At each gauss-point, the system was equilibrated for 0.5~ns and data collected over the subsequent
0.5~ns. The long-range contribution was obtained by performing particle insertion calculations in the appropriate molecular-shaped cavity (cf.~\ref{fg:cycle}) to  calculate $P(\varepsilon|\phi_\lambda)$ ($\lambda = 5$~{\AA}). Water with the appropriate
cavity was simulated for 1~ns and 1250 frames from the last 0.625~ns used for analysis. Confirming the gaussian-distribution of binding energies, particle extraction calculations agree with the particle insertion procedure (data not shown).

The excess energy was obtained by adapting the shell-wise calculation procedure used earlier for studying the hydration of methane \cite{asthagiri:jcp2008}. For the peptides, water molecules in the first two hydration shells were considered, where the hydration shell is defined by the union of shells of radius $\lambda$ centered on the heavy atoms. For ease, $\lambda \leq 5$~{\AA} defined the first shell, $5.0 < \lambda \leq 8$~{\AA} defined the second shell, and $8.0 < \lambda \leq 11.0$~{\AA}
defined the third shell. We find that the excess energy contributions from the third shell is close to zero, justifying our use of just the first two shells for calculating $h^{\rm ex}$. For calculating the excess energy we equilibrated the solvated peptide system an additional 1.5~ns (beyond what was used in the free energy calculation), and the propagated the trajectory for an additional 3~ns, collecting data every 500~ps for a total of 6000 frames. Entropies obtained using Eq.~\ref{eq:entropy} agree with entropy
calculated using $-\partial \mu^{\rm ex}/\partial T$ (data not shown). Further details on the entropy calculations will be described separately. 

In experiments, conditional or small molecule thermodynamic quantities are used in reconstructing the free energy of a macromolecule by scaling these quantities by the solvent-exposure in the macromolecule. We follow this procedure  and find solvent accessible surface area (SASA) using 
a standard code \cite{msms} and Bondi radii \cite{Bondi:1964jpc}. The area-scaled group-additive estimate of the free energy of the solute 
$S$ is given by 
\begin{eqnarray}
\mu^{\rm ex}_{S} & = & \alpha_{sc} \cdot \mu^{\rm ex}_{\rm sc-model}  +  \alpha_{back} \cdot \mu^{\rm ex}_ {\rm ref} \, ,
\label{eq:areascaling}
\end{eqnarray}
where $\mu^{\rm ex}_{\rm sc-model}$ refers to the side-chain model (either $g$-butane or isoleucine in GGIGG), $\mu^{\rm ex}_{\rm ref}$ is the hydration free energy of the reference for the remaining peptide, and 
$\alpha_{sc}$ and $\alpha_{back}$ are, respectively, the fractional solvent exposure of the side-chain 
and of the reference in $S$.  As an example of this procedure, to find the group-additive
free energy of the G$_9$I helix using the conditional hydration of isoleucine in the GGIGG peptide, we compute the SASA of G$_9$I, $A_t$, and identify
the contribution due to the side-chain alone, $A_{sc}$. The contribution of the rest of the peptide is then $A_{back} = A_t - A_{sc}$. We likewise find the
SASA of isoleucine in the reference GGIGG, $A_{sc}^{0}$, and of the G$_9$G peptide, $A_{back}^0$. Then the group-additive free energy
of G$_9$I is given by  $\mu^{\rm ex}[ {\rm G_9I}]  =  \alpha_{sc} \cdot \mu^{\rm ex} [{\rm I} | {\rm GGIGG}]  +  \alpha_{back} \cdot \mu^{\rm ex} [{\rm G_9G}]$,
with $\alpha_{sc} = {A_{sc}}/{A_{sc}^0}$ and $\alpha_{back} = {A_{back}} / {A_{back}^0}$.

\section{Results and discussion}

For an additive model to be satisfactory in predicting the thermodynamics of folding, Dill\cite{Dill:1997tg} has suggested that the
allowable error per group for a modest 100 amino-acid protein is about 10 cal/mol/group (or about 0.02~$k_{\rm B}T$/group at 298~K). This order of magnitude estimate sets the scale against which we can compare the success or failure of the 
additive approach. Beyond the numerical prediction of thermodynamic quantities, our aim is also to question if the physical conclusions based on translating small-molecule hydration data can be used to understand the hydration of a macromolecule.

\subsection{Hydration of isoleucine in pentapeptides}

Table~\ref{tb:penta} summarizes the results on the conditional hydration of isoleucine in the GGIGG and GGGGG pentapeptides. For reference we also
show the value expected from $g$-butane. 
\begin{table*}[h!]
\caption{Hydration thermodynamics of $g$-butane, GGGGG, and GGIGG. $k_{\rm B}T \ln x$ is the chemical contribution and reports on the consequences
of the attractive interaction between the solute and the solvent in the first hydration or inner-shell, here defined by the envelope obtained by the union of 
of 5~{\AA} spheres centered on the heavy atoms. $-k_{\rm B}T \ln p$ is the packing contribution measuring the work done to carve out the inner-shell
in the absence of the solute. $\mu^{\rm ex}_{\rm outer}$ is the contribution from solute interactions with the solvent outside the inner-shell when the
inner-shell is emptied of solvent.  $\mu^{\rm ex}$ is the net chemical potential, and $h^{\rm ex}$ and $Ts^{\rm ex}$ are its enthalpic and entropic 
components. $\alpha_{sc} = 0.619$ is the ratio of the solvent accessible surface area of the isoleucine side-chain in GGIGG to that for $g$-butane; $\alpha_{sc} = 0.616$
for IGGGG. All thermodynamic quantities are in units of kcal/mole.}\label{tb:penta}
\begin{tabular}{c | rrrrrr}
                         &   $k_{\rm B}T \ln x$   & $-k_{\rm B}T \ln p$ & $\mu^{\rm ex}_{\rm outer}$ & $\mu^{\rm ex}$ & $h^{\rm ex}$ & $Ts^{\rm ex}$ \\ \hline
$g$-Butane    & $-16.4\pm0.1$  & $24.0\pm0.1$ & $-5.1\pm 0.0 $ & $2.5\pm 0.1$ & $-3.4\pm 0.5$ & $-5.9 $ \\
GGGGG & $-85.8\pm0.2$ & $71.1\pm 0.2$ & $-17.3\pm0.1$ & $-32.0\pm 0.3$ & $-56.9\pm 1.4$ & $-24.8$ \\
GGIGG & $-88.2\pm0.2$ & $76.8\pm 0.2$ & $-17.7\pm 0.1$ & $-29.1\pm 0.3$ & $-56.3\pm 1.4$ & $-27.3$ \\ 
IGGGG & $-88.0\pm 0.2$ & $76.3\pm 0.2$ & $-18.3\pm 0.1$ & $-30.0\pm 0.3$ & $-57.3\pm 1.8$ & $-27.3$ \\  \hline
$\Delta$[GGIGG-GGGGG] & $\textbf{-2.4}\pm0.3$ & $\textbf{5.7}\pm 0.3$ & $\textbf{-0.4}\pm 0.1$ & $\textbf{2.9}\pm0.4$ & $0.6\pm 2.0$ & $-2.5$  \\
$\alpha_{sc}\cdot$[$g$-butane] & $\textbf{-10.2}$ & $\textbf{14.9}$ & $\textbf{-3.2}$ & $\textbf{1.5}$ & $-2.1$ & $-3.7$ \\ \hline
$\Delta$[IGGGG-GGGGG] & $\textbf{-2.2}\pm 0.3$ & $\textbf{5.2}\pm 0.3$ & $\textbf{-0.9}\pm 0.1$ & $\textbf{2.1}\pm 0.4$ & $-0.5\pm 2.0$ & $-2.6$  \\
$\alpha_{sc}\cdot$[$g$-butane] & $\textbf{-10.1}$ & $\textbf{14.8}$ & $\textbf{-3.1}$ & $\textbf{1.5}$ & $-2.1$ & $-3.7$ \\ \hline
\end{tabular}
\end{table*}

First note that the hydration thermodynamics of GGIGG and IGGGG are different. Thus it is not surprising that the conditional free energy of isoleucine will depend on where it is placed along the backbone. But it is the comparison of the conditional 
hydration of isoleucine in GGIGG with the hydration of $g$-butane that is revealing. As the above qualitative analysis 
(Sec.~\ref{sec:qual}) suggests, the packing contribution of  $g$-butane is substantially more positive than the corresponding conditional contribution. It is also apparent that (on a $k_{\rm B}T$-scale) the disagreement is disconcertingly large even after scaling for the fractional solvent exposure of isoleucine in the pentapeptide. Likewise, the chemistry contribution for $g$-butane is 
substantially more negative than the corresponding conditional contribution. Notice that the errors in the
packing and chemistry contributions tend to balance each other.  Lastly, the long-range contribution from $g$-butane
is also inutile in estimating the conditional long-range contribution of isoleucine.  

In the balancing of the various contributions noted above, one finds that the deviation in the net free energy is smaller: 
2.9~kcal/mol versus 1.5~kcal/mol for GGIGG and 2.1~kcal/mol versus 1.5~kcal/mol for IGGGG.  But in terms of the 
requirement for an additive model, the deviations are exceedingly large. 

The above conclusions are not altered for a different $\lambda$. The least credible definition of an inner-shell is one
that hugs the molecular surface. This is approximately the case for $\lambda = 3$~{\AA}. In this case, the chemistry contribution 
is zero (by definition), and the outer-contribution must be obtained from the invariance of $\mu^{\rm ex}$ with respect to $\lambda$ and the packing contribution at $\lambda = 3$~{\AA}. Using such estimates (data not shown), one still finds that the packing contribution for $g$-butane over-estimates the conditional packing contribution, and the long-range contribution of $g$-butane underestimates the corresponding conditional long-range contribution. 

Comparing $h^{\rm ex}$ and $Ts^{\rm ex}$, quantities that do not depend on $\lambda$, also indicates large
deviations between the conditional contribution and the area-scaled contribution from $g$-butane. Both the excess
entropy and the excess enthalpy of hydration of $g$-butane are more negative than the corresponding conditional contribution. In effect, using $g$-butane to model the isoleucine side-chain will lead one to ascribe a greater  hydrophobicity 
to the isoleucine side-chain than is warranted. Using somewhat different arguments, 
Ben-Naim \cite{BenNaimVol2} has come to the same conclusion. 

\subsection{Hydration of isoleucine in helical peptides}

Table~\ref{tb:helix} summarizes the results on the conditional hydration of isoleucine in the helical peptides G$_9$I and A$_9$I. 
\begin{table*}
\caption{Conditional hydration of isoleucine in deca-peptide models. (Gly)$_{10}$ (indicated as G$_9$G), (Gly)$_9\cdot$Ile (indicated as G$_9$I),  (Ala)$_{9}$Gly (indicated as A$_9$G), and (Ala)$_9\cdot$Ile (indicated as A$_9$I) are in the helical conformation. Substitutions with isoleucine are in position 6. The fractional solvent exposure of isoleucine in G$_9$I is $\alpha_{sc} = 0.59$ and in A$_9$I it is $\alpha_{sc} = 0.56$. Relative to isoleucine in GGIGG, $\alpha_{sc} = 0.95$ (G$_9$I) and
$\alpha_{sc} = 0.90$ (A$_9$I). Rest as in Table~\ref{tb:penta}.}\label{tb:helix}
\begin{tabular}{c | rrrr}
                         &   $k_{\rm B}T \ln x$   & $-k_{\rm B}T \ln p$ & $\mu^{\rm ex}_{\rm outer}$ & $\mu^{\rm ex}$ \\ \hline
G$_9$G & $-95.1\pm 0.3$ & $78.0\pm 0.7$ & $-32.2\pm 0.2$ & $-49.3$  \\
G$_9$I & $-97.3\pm 0.2$ & $83.7\pm 0.5$ & $-32.0\pm 0.6$ & $-45.6$ \\ \hline
$\Delta$[G$_9$I - G$_9$G] &  $\textbf{-2.2}\pm 0.4$ & $\textbf{5.7}\pm 0.9$ & $\textbf{0.2}\pm 0.6$ & $\textbf{3.7}\pm 1.2$  \\
$\alpha_{sc}\cdot$[$g$-butane] & $\textbf{-9.7}$ & $\textbf{14.2}$ & $\textbf{-3.0}$ & $\textbf{1.5}$  \\ 
$\alpha_{sc}\cdot$[GGIGG $-$ GGGGG] & $\textbf{-2.3}$ & $\textbf{5.4}$ & $\textbf{-0.4}$ & $\textbf{2.7}$  \\ \hline
A$_9$G & $-95.1\pm 0.2$ & $86.7\pm 0.7$ & $-30.2\pm 0.5$ & $-38.6$  \\
A$_9$I & $-95.8\pm 0.2$ & $90.7\pm 0.6$ & $-30.3\pm 0.5$ & $-35.4$ \\ \hline
$\Delta$[A$_9$I - A$_9$G] &  $\textbf{-0.7}\pm 0.3$ & $\textbf{4.0}\pm 0.9$ & $\textbf{-0.1}\pm 0.7$ & $\textbf{3.2}\pm 1.2$  \\
$\alpha_{sc}\cdot$[$g$-butane] & $\textbf{-9.2}$ & $\textbf{13.4}$ & $\textbf{-2.9}$ & $\textbf{1.3}$  \\ 
$\alpha_{sc}\cdot$[GGIGG$-$GGGGG] & $\textbf{-2.2}$ & $\textbf{5.1}$ & $\textbf{-0.4}$ & $\textbf{2.5}$  \\ \hline
\end{tabular}
\end{table*}   
As was found for the pentapeptide system, the free energy components for $g$-butane are substantially 
different from the conditional quantities, with the local chemical contribution being significantly underestimated and the 
packing contribution being significantly overesimated. 

Comparing G$_9$G and A$_9$G clearly highlights the importance of the background in assessing hydration thermodynamics of amino acid side-chains. The G$_9$G peptide is more polar than A$_9$G and hence the work required to open a cavity near the G$_9$G helix face to accommodate the isoleucine is nearly 2~kcal/mol higher than that in A$_9$G. For the same reason, clustering of water molecules is more favorable around the side-chain in the G$_9$G framework than in A$_9$G. 
Not surprisingly, these trends largely balance and the net conditional free energy of isoleucine, $3.7$~kcal/mol in G$_9$I versus
$3.2$~kcal/mol inA$_9$I, is different by only 0.5~kcal/mol.

As mentioned earlier, oftentimes in experimental analysis the conditional hydration of the residue X in a GXG peptide
is used to model the hydration of the residue in other contexts. We follow the same procedure here and compare
the conditional hydration of isoleucine in the GGIGG model to the conditional hydration of isoleucine in the helices. 
We also correct for the fractional exposure of the side-chain in the helical peptides relative to that in the GGIGG model. 
In this instance, the qualitative arguments developed above do not apply directly, since hydration of the side-chain
model is not referenced relative to pure solvent. Nevertheless, in terms of the requirement imposed on an additive model, 
it is again apparent that the conditional hydration of isoleucine in GGIGG deviates substantially from that in the helical peptides. 

Finally, just as experimentalist do for real proteins, we attempted to reconstruct the free energy of the helical peptides using the conditional hydration free energy of isoleucine in the GGIGG model. Table~\ref{tb:hrconst} summarizes the results of this 
exercise. 
\begin{table*}
\caption{Predicted thermodynamics of G$_9$I and A$_9$I helical peptides using Eq.~\ref{eq:areascaling}, with the conditional hydration contribution of isoleucine modeling the side-chain and G$_9$G or A$_9$G the reference peptide. $\alpha_{sc}$ is the fractional solvent exposure of the side-chain in the helical system relative to its exposure in GGIGG. $\alpha_{back}$ is fractional exposure of the non-isoleucine parts of G$_9$I or A$_9$I
relative to the appropriate reference. $\Delta$ is the error in the predicted value relative to the simulated value. The simulated values of G$_9$I and A$_9$I are reproduced from Table~\ref{tb:helix}. All values are in kcal/mole.}\label{tb:hrconst}
\begin{tabular}{c | rccr}
                                &   $k_{\rm B}T \ln x$   & $-k_{\rm B}T \ln p$ & $\mu^{\rm ex}_{\rm outer}$ & $\mu^{\rm ex}$ \\ \hline
G$_9$I                   & $-97.3$         & $83.7$         & $-32.0$                      & $-45.6$ \\ 
${\rm [G_9I]}_{\rm A}$  & $-90.3$          & $77.6$         & $-30.2$                      & $-42.9$ \\
$\Delta$ &  $\textbf{7.0}$ & $\textbf{-6.1}$ & $\textbf{1.8}$ & $\textbf{2.8}$ \\ \hline
A$_9$I                             & $-95.8$ & $90.7$ & $-30.3$ & $-35.4$ \\ 
${\rm [A_9I]}_{\rm A}$ & $-91.5$ & $86.5$ & $-28.7$ & $-33.7$ \\
$\Delta$ & $\textbf{4.3}$ & $\textbf{-4.2}$ & $\textbf{1.6}$ & $\textbf{1.7}$ \\ \hline
\end{tabular}
\end{table*}
It is evident that the reconstructed free energy of the helices deviate from the simulated values by at least 1.5~kcal/mole, and the small net deviation results from large compensating deviations in the predicted chemical, packing, and long-range contribution. For both G$_9$I and A$_9$I,  the magnitude of the minimum error is set by the errors in the long-range contribution to hydration. As suggested in Sec.~\ref{sec:qual}, trying to model the long-range contributions using 
a scaling based on surface areas will have limitations: the surface area scaling is effectively a local construct, whereas the long-range interactions, as the name indicates, involves interaction of the group with the entire solvent bath.

The above analysis shows that one of the most often used premises to model and/or interpret protein hydration, namely using data from GXG after scaling for solvent-exposure, is inadequate in predicting the hydration of the same residue in a somewhat different context. Therefore, all results based on such constructions must be considered with caution.

\section{Concluding discussions}

There are two main conclusions that emerge from this work. First, the conditional solvation of an amino acid in a peptide model is rather different from the solvation of the corresponding amino acid side-chain in bulk solvent. This point is has
been made before, most notably by Ben-Naim. What we show here is that assuming a non-polar solute ($g$-butane) models the side-chain of isoleucine one would ascribe to the isoleucine residue a greater hydrophobicity than is warranted. We anticipate that even the temperature dependencies of the  conditional packing contribution to be rather different from the packing contribution for the nonpolar solute.  On this basis concerns are raised about drawing conclusions about the presumed dominance of hydrophobic hydration in protein folding based on small molecule hydration. 

The second allied conclusion  is that even if an additive description of free energy predicts the numerical values in 
reasonable accord with experiments, it is likely that this success  arises primarily due to compensating failures in 
describing the underlying physics.

\begin{acknowledgements}
We thank Dr.~Gillian Lynch for providing the structure of a deca-alanine in the helical conformation. We thank Arieh Ben-Naim for helpful comments on the manuscript. BMP gratefully acknowledge the financial support of the National Institutes of Health (GM 037657),  the National Science Foundation (CHE-1152876) and the Robert A. Welch Foundation (H-0037). This research used resources of the National Energy Research Scientific Computing Center, which is supported by the Office of Science of the U.S. Department of Energy under Contract No. DE- AC02-05CH11231.
\end{acknowledgements}


\end{document}